\documentclass[preprint,pre,showpacs]{revtex4}
\usepackage[english]{babel}
\usepackage{graphicx}
\usepackage{graphics}
\usepackage{epsfig}
\usepackage{multirow}
\usepackage{amsmath}
\usepackage{amssymb}
\usepackage{amsfonts}

\begin{document}
\title{Inhomogeneities on all scales at a phase transition altered by disorder}
\author{I. Balog}
\email{balog@ifs.hr}
\author{K. Uzelac}
\email{katarina@ifs.hr}
\affiliation{Institute of Physics, P.O.Box 304, Bijeni\v{c}ka cesta 46, HR-10001 Zagreb, Croatia}
\begin{abstract}
   We have done a finite-size scaling study of a continuous phase transition altered by the quenched bond disorder, investigating systems at quasicritical temperatures of each disorder realization by using the equilibriumlike invaded cluster algorithm. Our results indicate that in order to access the thermal critical exponent $y_\tau$, it is necessary to average the free energy at quasicritical temperatures of each disorder configuration. Despite the thermal fluctuations on the scale of the system at the transition point, we find that spatial inhomogeneities form in the system and become more pronounced as the size of the system increases. This leads to different exponents describing rescaling of the fluctuations of observables in disorder and thermodynamic ensembles.  
\end{abstract}
\pacs{05.50.+q, 64.60.F-, 75.10.Hk, 2.70.-c}
\maketitle
   The question of how disorder modifies a phase transition is still not fully resolved, notwithstanding decades of research. Many issues remain open, even in the case of the ferromagnetic transition in the presence of quenched disorder, from the qualitative nature of the ordered phase near criticality \cite{TarjDots_06,KRSZ_2011}, the presence of Griffiths singularities \cite{Griff}, the role of multiple length scales \cite{Korzvn}, to violation \cite{PazScaZi} of the proposed \cite{CCSdsF} bound on the thermal critical exponent.

   An important aspect present in the majority of the cited problems is the non-self-averaging effect between different disorder realizations \cite{WiseDom_95}, leading to the question of proper averaging over disorder. The standard approach is to reduce the problem to an effective translationally invariant one, by averaging the free energy over disorder configurations at a given temperature \cite{GrnLuth}. Such averaging permits one to apply the perturbative renormalization group (RG), and crucially simplifies the finite-size scaling (FSS) studies based on numerical simulations as well. When applied to a classical ferromagnetic model with bond randomness, displaying a continuous phase transition in the pure case such as the one we consider in this paper, the conclusion seems to be \cite{Vojta06} that the disorder, if relevant in the sense of the Harris criterion \cite{Harris}, merely changes the values of the critical exponents, while the nature of the transition remains similar to that of the pure case. 

   We challenge this notion with a FSS study in which the average is taken at quasicritical (i.e. finite size critical) temperatures of each disorder sample. Except for some partial attempts, a systematic study using this type of averaging is still missing, since it requires an exceptional numerical effort. The recently proposed equilibriumlike invaded cluster (EIC) algorithm \cite{BU_1,BU_2} enables such a procedure to be technically feasible. 

   There exist prior works revealing conceptual problems in the transition with bond disorder. It has been pointed out that the self-averaging ratios \cite{WiseDom_95} of observables at criticality depend on averaging \cite{WiseDom} although they should be universal \cite{AhaHarWis}. Correlation length has also been shown to be non-self-averaging near criticality \cite{PPS04}. P\'azm\'andi \textit{et. al.} \cite{PazScaZi} have shown that standard averaging obscures the intrinsic thermal exponent $y_\tau=1/\nu$, when it is superior to the exponent $\tilde{y}_\tau$ describing disorder fluctuations of quasicritical temperatures,
\begin{equation}
\label{Pazm}
\tilde{y}_\tau<y_\tau.
\end{equation}
In such a case, in standard averaging, $\tilde{y}_\tau$ always takes over $y_\tau$ in FSS relations, and the information about $y_\tau$ is lost. It has therefore been suggested \cite{WiseDom,BerPazBat} that in order to find the true critical exponents one has to perform a disorder average of observables, taken at the quasicritical temperature of each disorder sample.

   We use the EIC approach on the two-dimensional (2D), three-state Potts model with bond disorder \cite{rev_rndPtts}, where disorder is relevant by the Harris criterion, since the specific-heat critical exponent of the pure model is positive ($\alpha=1/3$). Our main result shows inequality (\ref{Pazm}) by an explicit numerical calculation, justifying the assumption by P\'azm\'andi \textit{et al.} \cite{PazScaZi} in a classical system. We suggest an alternative interpretation of the lack of self-averaging as an emergence of frozen inhomogeneities in observables at all scales, which may be analyzed further from the  statistics of the largest cluster at criticality. 

   Since the Chayes bound \cite{CCSdsF}, i.e. $\tilde{y}_\tau\leq d/2$, is exclusively a result of averaging at a unique temperature, it is irrelevant for the intrinsic thermal exponent $y_\tau$. For this reason our findings might also be of interest to systems in which a violation of the Chayes criterion was observed \cite{viol_1,viol_1t5,viol_2}. Experimentally, the intrinsic exponent $y_\tau$ would be observed only in direct calorimetric measurements of the heat capacity, while the indirect measurements (e.g. birefringence \cite{Birg2}) would yield the exponent $\tilde{y}_\tau$ obeying the Chayes criterion.

   We introduce disorder in the Potts model \cite{Ptts} as a random bond dilution  
\begin{equation}
\label{dil_Potts}
H=\sum_{<i,j>}-J_{i,j}\big(\delta_{\sigma_i,\sigma_j}-1\big),
\end{equation}
\noindent where random couplings $J_{i,j}$ are restricted to neighboring lattice sites $i$, $j$ and take zero value with concentration $\boldsymbol{c}$, or $J_{i,j}=J$ otherwise. $\sigma$ denotes the Potts variable with $q=3$ discrete states. For the purpose of numerical simulations, the partition function of the Potts model is written in terms of the random cluster model, by using the Fortuin and Kasteleyn (FK) graph expansion \cite{FK} 
\begin{equation}
\label{rand_cluster}
Z=\sum_{\gamma \in\Gamma_{\boldsymbol{\alpha}}}p^{b(\gamma)} \cdot (1-p)^{B-b(\gamma)} \cdot q^{c(\gamma)},
\end{equation}
where the  bond probability
\begin{equation}
\label{p_def}
p=1-e^{-\frac{J}{k_BT}}
\end{equation} 
takes the role of temperature, while $b(\gamma)$ and $c(\gamma)$ denote the number of bonds and  connected components ``FK clusters" in the graph $\gamma$, respectively. The summation runs over the set of all possible FK graphs $\Gamma_{\boldsymbol{\alpha}}$, compatible with the given disorder configuration $\boldsymbol{\alpha}$.

   When applied to a system with disorder, the EIC algorithm \cite{BU_1} simulates it at the quasicritical (i.e., finite-size critical) bond probabilities $p^c_{\boldsymbol{\alpha}}(L)$ (related to temperature by Eq. (\ref{p_def})) belonging to each disorder configuration $\boldsymbol{\alpha}$, defined by the onset of the percolation of the largest FK cluster. This property allows us to consider separately the scaling of thermal fluctuations of an observable around its thermodynamic mean and the disorder ensemble fluctuations of the thermodynamic means. Our algorithm is an extension of the invaded cluster (IC) algorithm \cite{ic} and has a similar mechanism of self-regulating to the quasicritical point, but with the crucial difference of generating the equilibrium thermodynamic ensemble \cite{BU_2}. The duration of a Monte Carlo (MC) step of the EIC algorithm is approximately the same as that of the IC algorithm, but because of the canonical constraint, the EIC algorithm requires a certain number of thermalization steps (never exceeding $5000$ in the present paper). The EIC algorithm generalizes to the problem with bond dilution in a straightforward way since disorder merely excludes some configurations from the set of all possible FK graphs of the pure case. 

   Throughout this paper we denote by $[\cdot]$ the disorder average taken at $p^c_{\boldsymbol{\alpha}}(L)$ of each disorder configuration and by $\overline{{}\cdot{}}$ thermodynamic average for a given $\boldsymbol{\alpha}$. Numerical results presented are based on simulations on square lattices of linear size $L$ ranging from $64$ up to $896$, with two disorder concentrations, $c=0.125$ and $0.25$, using the statistics of $400$ and $600$  disorder configurations respectively, with $20000$ Monte Carlo steps (MCS) per disorder configuration after thermalization. Calculation of a running average reveals that such a disorder statistics is sufficient to determine each $[p^c_{\boldsymbol{\alpha}}(L)]$ to seven significant digits. The disorder is introduced microcanonically, i.e. the exact number of bond vacancies corresponding to a concentration is randomly distributed on the lattice.

\begin{figure*}
\begin{center}
\begin{picture}(416,160)
\put(0,0){\includegraphics[width=416pt,height=150pt]{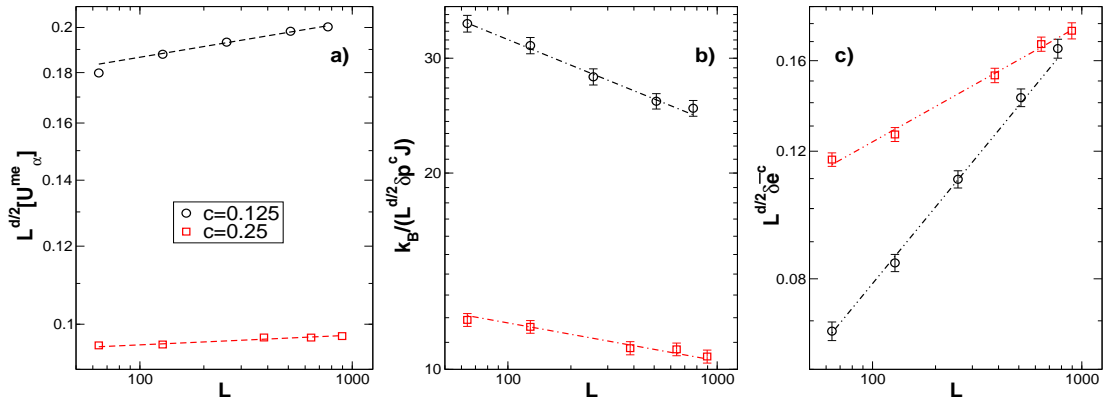}}
\end{picture}
\caption{\footnotesize{(color online) Slopes in (a)-(c) correspond to $y_\tau-\frac{d}{2}$, $\tilde{y}_\tau-\frac{d}{2}$ and $\frac{d}{2}-\tilde{x}$ respectively, pointing out the difference between the three exponents. The disorder statistics determines the precision of $[U^{me}_{\boldsymbol{\alpha}}]$ for each $L$ to the fourth significant digit and the precision of $\delta p^c$ and $\delta\overline{e}^c$ to the third significant digit. The error bars for $L^{d/2}[U^{me}_{\boldsymbol{\alpha}}]$ are smaller than the symbols.}}
\label{f5}
\end{center}
\end{figure*}

    The thermal critical exponent $y_\tau$ is calculated from the magnetization-energy cumulant  
\begin{equation}
\label{me}
U^{me}_{\boldsymbol{\alpha}}=\frac{\overline{m_{\boldsymbol{\alpha}}e_{\boldsymbol{\alpha}}}-\overline{m_{\boldsymbol{\alpha}}}\cdot\overline{e_{\boldsymbol{\alpha}}}}{\overline{m_{\boldsymbol{\alpha}}}},
\end{equation}
where $m_{\boldsymbol{\alpha}}$ and $e_{\boldsymbol{\alpha}}$ denote the order parameter and energy density of a disorder configuration $\boldsymbol{\alpha}$, respectively. The values of $y_\tau$ in Table \ref{t1} are obtained by fitting (Fig. \ref{f5} a) the averaged data to the power law form $[U^{me}_{\boldsymbol{\alpha}}] \propto L^{y_\tau-\boldsymbol{d}}$. The simple power law describes the scaling of $[U^{me}_{\boldsymbol{\alpha}}]$ for $c=0.25$ in the entire range of lattice sizes and for $c=0.125$ only the data from the smallest size deviate \cite{BU_nered2}. Since the scaling corrections are always found to be important in approaches using the averaging at a unique temperature (see e.g. \cite{hasen}), their negligibility can only be attributed to the averaging procedure we used in this work. Exponent $y_\tau$ also shows a negligible concentration dependence, contrary to the previous studies where the standard averaging was used \cite{Balestero}. 
\begin{table}[!h]
\caption{\footnotesize{Rescaling exponents of $[U^{me}_{\boldsymbol{\alpha}}]$, $\delta p^c_{\boldsymbol{\alpha}}$ and $\delta\overline{e}^c_{\boldsymbol{\alpha}}$. Each set of data has been fitted in the entire range of sizes to a single power-law except for the $[U^{me}_{\boldsymbol{\alpha}}]$ for $c=0.125$, where the smallest lattice has been excluded. Error bars have been estimated by jackknife binning.
}}
\centering{
\footnotesize{\begin{tabular}{c|c|cc}
\hline
c & $y_\tau$ & $\tilde{y_\tau}$ & $d-\tilde{x}$ \\
\hline
$0.125$ & $\boldsymbol{1.03(1)}$ & $0.88(4)$ & $1.35(5)$ \\
$0.25$ & $\boldsymbol{1.01(1)}$ & $0.95(2)$ & $1.14(4)$ \\
\hline
\end{tabular}}
}
\label{t1}
\end{table} 

  To understand how the exponent $y_\tau$ becomes obscured when the averaging at a unique temperature is used, we examine the fluctuations of quasicritical bond probabilities $p^c_{\boldsymbol{\alpha}}$ (Fig. \ref{f5} b) in disorder ensemble (ensemble of random samples)
\begin{equation}
\label{pscrt} 
\delta p^c  = \sqrt{[p^{c 2}_{\boldsymbol{\alpha}}]-[p^c_{\boldsymbol{\alpha}}]^2}\propto L^{-\tilde{y}_\tau}.
\end{equation}
By virtue of Eq. (\ref{p_def}) $p^c$ obeys the same scaling law as the fluctuations of quasicritical temperatures $T^c_{\boldsymbol{\alpha}}$. The fact that beyond statistical errors $\tilde{y}_\tau$ is dominant over $y_\tau$ (Tab. \ref{t1}), leads to the conclusion that, by averaging at a unique temperature one effectively measures the exponent $\tilde{y}_\tau$ \cite{PazScaZi}. The range of values for different disorder concentrations that we have obtained for $\tilde{y}_\tau$ corresponds to the findings for the thermal exponent in previous studies, which used averaging at a unique temperature \cite{CardJac,Ludw,DotPicPuy} for the same system. For example Jacobsen and Cardy \cite{CardJac} have estimated the value to $0.96(4)$. Unlike $y_\tau$, the exponent $\tilde{y}_\tau$ displays a strong dependence on disorder, similar to that found in previous studies. 

  From $U^{me}_{\boldsymbol{\alpha}}$ we have determined the singular part of the energy density which scales $\propto L^{y_\tau-d}$. Considering disorder ensemble fluctuations of thermodynamic means of energy density $\overline{e}^c_{\boldsymbol{\alpha}}$ taken at $T^c_{\boldsymbol{\alpha}}$ (Fig. \ref{f5} c),
\begin{equation}
\label{Eqscrt}
\delta\overline{e}^c = \sqrt{[\overline{e}^{c 2}_{\boldsymbol{\alpha}}]-[\overline{e}^c_{\boldsymbol{\alpha}}]^2}\propto L^{-\tilde{x}},
\end{equation}
we obtain that $L^d \delta\overline{e}^c$ increases, with the exponent $d-\tilde{x}$ (Tab. \ref{t1}) significantly larger than ${y}_\tau$ or $\tilde{y}_\tau$ for both $c$. Since $\delta\overline{e}^c$ is the sum of the analytic and the singular part, the exponent $\tilde{x}$ has to be attributed to the analytic part, which is dominant over the singular one. In Ref. \cite{CardJac} the analysis of the energy fluctuations by the standard disorder averaging produces the exponent $d-0.91(1)=1.09(1)$, significantly larger than their value for the thermal exponent $0.96(4)$. 

   If the disorder had merely changed the values of the critical exponents found for the pure case, $\tilde{y}_\tau$ and $d-\tilde{x}$ would have to be equal to $y_\tau$, as determined by the singular part of the free energy. However, we show the exponents to be different (Fig. \ref{f5}).


   The evidence presented leads to the conclusion that changing a disorder configuration $\boldsymbol{\alpha}$ while keeping the concentration of disorder constant, drives the system out of the critical area. A consequence of this scenario is the formation of spatial inhomogeneities at all scales. Consider the free energy density for a single disorder configuration $\boldsymbol{\alpha}$
\begin{equation}
\label{hom_nered}
f_{\boldsymbol{\alpha}}=L^{-d}\hat{f}_{\boldsymbol{\alpha}}(L^{y_{\tau}}(T-T^c_{\boldsymbol{\alpha}}(L)),L^{y_h}h)+T\cdot\tilde{e}_{\boldsymbol{\alpha}}(T,h;L),
\end{equation} 
\noindent where $h$ is the magnetic field and $T\tilde{e}_{\boldsymbol{\alpha}}$ is the analytical part. Since we have demonstrated that inequality (\ref{Pazm}) applies, changing $\boldsymbol{\alpha}$ produces a shift in the argument of expression (\ref{hom_nered}) $\propto L^{y_\tau-\tilde{y}_ \tau}$, which diverges for $L\to\infty$. However, when a system of size $L$ can be driven out of criticality by changing $\boldsymbol{\alpha}$, this shift is even larger for any of its parts of size $L'<L$ since $L'^{-\tilde{y}_\tau}>L^{-\tilde{y}_\tau}$. The shift may be positive or negative in any of the parts since the disorder configurations in them are uncorrelated, so the system locally prefers either the low- or the high- temperature phase, resulting in spatial inhomogeneities. The contributions from local parts accumulate in the analytic part of the energy, causing the disorder ensemble fluctuations (i.e., the exponent $\tilde{x}$). If the opposite to inequality (\ref{Pazm}) were true, the inhomogeneities would have disappeared in the limit $L\to\infty$, since $L^{y_\tau-\tilde{y}_ \tau}$ would be at most of order $1$.
 
\begin{figure}
\begin{center}
\begin{picture}(180,184)
\put(0,0){\includegraphics[width=180pt,height=184pt]{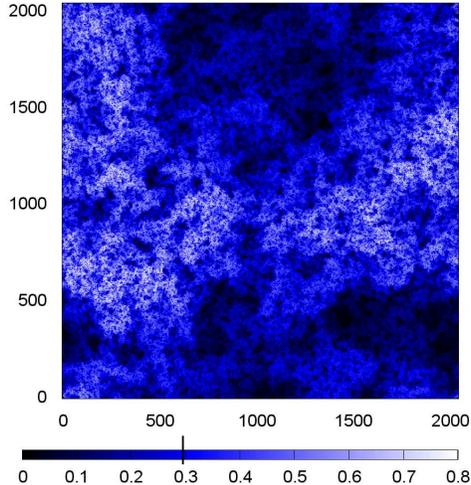}}
\put(68,10){$\boldsymbol{|}$}
\end{picture}
\caption{\footnotesize{(Color online) Spatial dependence of $\overline{m}_{\boldsymbol{\alpha}}(\vec{r})$ for a single disorder configuration at $p^c_{\alpha}$ obtained with $10^6$ MCS for $L=2048$. The spatial average of  $\overline{m}_{\boldsymbol{\alpha}}(\vec{r})$ in this disorder configuration is $0.29164$ and is labeled by the tick on the scale.}}
\label{f1}
\end{center}
\end{figure}

   We demonstrate the existence of the spatial inhomogeneities on the example of the local order parameter $\overline{m}_{\boldsymbol{\alpha}}(\vec{r})$, defined as the thermodynamic probability that the spin on the site $\vec{r}$ belongs to the largest FK cluster
\begin{equation}
\label{lok_m}
\overline{m}_{\boldsymbol{\alpha}}(\vec{r})=\frac{\sum_{\{\sigma\},\{b\}}\pi_{\boldsymbol{\alpha}}(\vec{r})\cdot w_{\boldsymbol{\alpha}}(\{\sigma\},\{b\})}{\sum_{\{\sigma\},\{b\}}w_{\boldsymbol{\alpha}}(\{\sigma\},\{b\})},
\end{equation}
\noindent where $\pi_{\boldsymbol{\alpha}}$ is $1$ if $\sigma(\vec{r})$ belongs to the largest FK cluster and $0$ otherwise. By $w_{\boldsymbol{\alpha}}$ we denote the statistical weight of a configuration in the joint space of spin and bond variables. The spatial average of $\overline{m}_{\boldsymbol{\alpha}}(\vec{r})$ is equivalent to the standard Potts order parameter \cite{StrMEF}. In Fig. \ref{f1} $\overline{m}_{\boldsymbol{\alpha}}(\vec{r})$ in a single disorder sample with $25\%$ of disorder at $p^c_{\boldsymbol{\alpha}}(L)$ is presented. The behavior of $\overline{m}_{\boldsymbol{\alpha}}(\vec{r})$ contrasts with the situation in the pure model, where all the lattice sites are equivalent and $\overline{m}_{\boldsymbol{\alpha}}(\vec{r})$ is constant. It illustrates how a fractal volume, pinned by disorder, emerges as the preferred ordering volume on which the local order parameter is close to $1$. 
\begin{figure}
\begin{center}
\begin{picture}(160,120)
\put(0,0){\includegraphics[width=160pt,height=120pt]{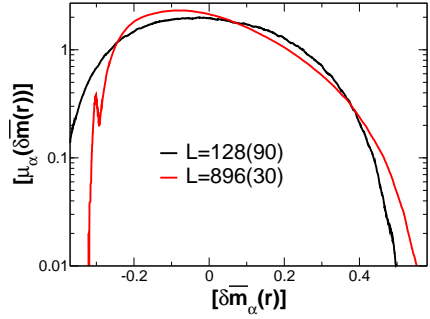}}
\end{picture}
\caption{\footnotesize{(color online) Distributions of spatial variations of the order parameter for two lattice sizes averaged over several disorder configurations (in parentheses) for $c=0.25$.}}
\label{f2}
\end{center}
\end{figure}
\begin{figure}
\begin{center}
\begin{picture}(160,160)
\put(0,0){\includegraphics[width=160pt,height=160pt]{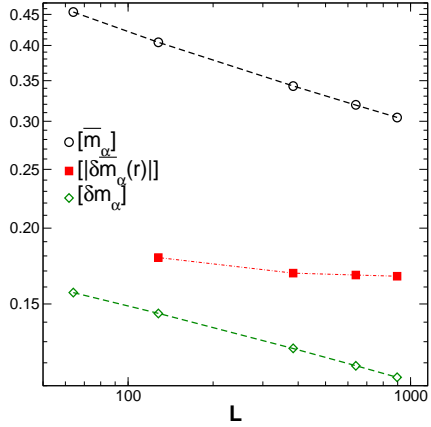}}
\end{picture}
\caption{\footnotesize{(Color online) Size dependence of: a) order parameter $[\overline{m}_{\boldsymbol{\alpha}}]$, b) width of the spatial variations of the local order parameter $[|\delta\overline{m}_{\boldsymbol{\alpha}}(\vec{r})|]$, and c) width of the thermal fluctuations of the order parameter $[\delta m_{\boldsymbol{\alpha}}]$, for $c=0.25$. The dashed lines denote the best fits which include scaling corrections.}}
\label{f3}
\end{center}
\end{figure}

   We define $\mu_{\boldsymbol{\alpha}}(\delta\overline{m}_{\boldsymbol{\alpha}})$ as the probability distribution of the deviations of the local order parameter from its spatial mean $\delta \overline{m}_{\boldsymbol{\alpha}}(\vec{r})=\overline{m}_{\boldsymbol{\alpha}}(\vec{r})-(1/V)\int\overline{m}_{\boldsymbol{\alpha}}(\vec{r})d\vec{r}$ at $p^c_{\boldsymbol{\alpha}}(L)$, for each $\boldsymbol{\alpha}$. A comparison of the averaged distributions $[\mu_{\boldsymbol{\alpha}}]$ for the two different lattice sizes (Fig. \ref{f2}) reveals that the tail of the distribution, related to the significant deviations from the average ($\delta \overline{m}_{\boldsymbol{\alpha}}\approx 1$), becomes more pronounced with increasing system size, although the order parameter vanishes in the limit $L\to\infty$. As shown in Fig. \ref{f3}, the width of the spatial inhomogeneities of order parameter $[|\delta\overline{m}_{\boldsymbol{\alpha}}(\vec{r})|]$ is larger than the width of thermal fluctuations of the order parameter $[\delta m_{\boldsymbol{\alpha}}]=\sqrt{[\overline{m^2}_{\boldsymbol{\alpha}}-\overline{m}^2_{\boldsymbol{\alpha}}]}$.  We conclude that spins which contribute to ordering are increasingly frozen as $L$ increases.  

   The magnetic exponent ($\frac{\beta}{\nu}$) calculated for $c=0.125$ and $0.25$ from the first moment of the order parameter ($0.129(3)$) which includes the contribution from inhomogeneities does not differ significantly from the value obtained from the second moment ($0.132(3)$), which includes only thermal fluctuations \cite{BU_nered2}. Both of these values are within the range of previous results for the exponent $\beta/\nu$ \cite{CardJac,Ludw,DotPicPuy}. We conclude that the difference in $\overline{m}_{\boldsymbol{\alpha}}$ between different disorder configurations $\boldsymbol{\alpha}$, when determined at $p^c_{\alpha}$, is of the same order as the width of the thermal fluctuations of the order parameter for each $\boldsymbol{\alpha}$. 

   In conclusion, we have used the recently proposed EIC algorithm to study the critical behavior in the $2D$ $q=3$ Potts model with quenched disorder by averaging at quasicritical temperatures of individual disorder configurations. This procedure has allowed us to separate the thermal from disorder-sample fluctuations and obtain the critical exponents $y_\tau$ and $y_h=d-\beta/\nu$ characterizing the transition in a single disorder configuration. The difference between the critical and disorder fluctuation exponents is interpreted by the emergence of spatial inhomogeneities on all scales at the transition point. By examining the local order parameter, we explicitly show that the inhomogeneities are present and that they become increasingly frozen (localized) as the size of the system increases. The above studies are currently being extended to higher dimensions, by using the same EIC algorithm. The approach may be easily applied to more specific problems, such as the extensively discussed 2D Ising model \cite{ggkr09} as a marginal case of quenched disorder according to the Harris criterion, or to the change from a first- to second-order phase transition induced by disorder \cite{rev_rndPtts}.

{\it Note added}. Another interesting extension of the application of the EIC algorithm would include models with random field, for which an alternative approach to eliminate the lack of self-averaging has been recently proposed \cite{FMY11}. 

\begin{acknowledgements}
I.B. wishes to thank O. S. Bari\v{s}i\'c for useful discussions and comments. This work was supported by the Croatian Ministry of Science, Education and Sports through Grant No. 035-0000000-3187. We thank the referee for driving our attention to Ref. \cite{FMY11}.
\end{acknowledgements}  

\end{document}